\documentclass{ws-ijmpa}

\usepackage{subfigure}
\begin{document}

\markboth{G D Lafferty} {TAU PHYSICS FROM B FACTORIES}

%
\catchline{}{}{}{}{}
%

\title{\hfill MAN/HEP/2006/28 \\ \hfill \\ TAU PHYSICS FROM B FACTORIES }

\author{G D LAFFERTY}

\address{School of Physics and Astronomy, The University of Manchester\\
Manchester M13 9PL, UK\\
George.Lafferty@manchester.ac.uk}

\maketitle

\begin{history}

Presented at Charm 2006, International Workshop on Tau-Charm
Physics, June 05-07 2006, Beijing, China
\end{history}

\begin{abstract}
Some recent $\tau$-physics results are presented from the BaBar
and Belle experiments at the SLAC and KEK B factories, which
produce copious numbers of $\tau$-lepton pairs. Measurements of
the tau mass and lifetime allow to test lepton universality and
CPT invariance, while searches for lepton-flavour violation in tau
decays are powerful ways to look for physics beyond the Standard
Model. In semihadronic, non-strange tau decays, the vector
hadronic final state is particularly important in helping
determine the hadronic corrections to the anomalous magnetic
moment of the muon, while studies of strange final states are the
best available ways to measure the CKM matrix element $V_{\rm us}$
and the mass of the strange quark. 
\end{abstract}

\ccode{PACS numbers: 13.35.Dx, 13.66.De, 14.60.Ef}

\section{Tau-Pair Production at B Factories}

The SLAC and KEK B factories are asymmetric e$^+$e$^-$ colliders
which run at centre-of-mass energies at, and close to, the
$\Upsilon(4{\rm S})$ resonance. At these energies, around
10.58~GeV, the cross section for tau-pair production, at 0.89~nb,
is close to the 1.05~nb cross section for B$\overline {\rm B}$
production. These machines are therefore also tau factories, and
the experiments BaBar and Belle have between them so far recorded
over $10^9$ tau-pair events. In comparison, the four LEP
experiments each recorded only about $10^5$ tau-pair events,
mainly running on the Z resonance, while CLEO-III at CESR took
about $7 \times 10^6$ tau-pair events. However, it should be noted
that experimental conditions for tau physics at LEP were
particularly clean, resulting in relatively small systematic
errors. For channels with large branching fractions, where
systematic errors dominate, the LEP results will remain
competitive for some time.

\section{Testing the Standard Model}

Measuring static properties of the tau, such as mass, lifetime and
electroweak couplings, allow tests of lepton universality and CPT
invariance. Searches for lepton-flavour violation (LFV) in tau
decays are particularly topical. In the Standard Model (SM),
minimally augmented with massive neutrinos and neutrino
oscillations, LFV decays of the tau are allowed, but at an
unobservably low level~\cite{ref:LFVSM}. However, various
extensions to the SM predict LFV rates accessible to the B
factories, with branching fractions of order $~10^{-8}$ and
higher~\cite{ref:LFVBSM}.

\subsection{Tests of CPT and lepton universality}

\begin{figure}[pb]
\centerline{\psfig{file=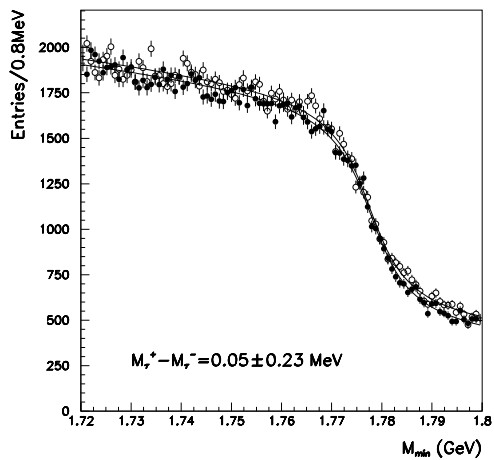,width=0.6\textwidth}}
\vspace*{8pt} \caption{The distributions of the pseudomass (from
Belle) for the decays $\tau^\pm \to 3\pi^\pm\nu_\tau$ shown
separately for $\tau^+$ (solid points) and $\tau^-$ (open points)
decays. The solid curves show fits to obtain the tau mass.
\label{fig:MassDiff}}
\end{figure}

Belle has used 3-prong tau decays from the process $\tau^- \to
\pi^-\pi^-\pi^0\nu_\tau$, in a sample of 414~fb$^{-1}$ of data, to
measure the tau mass and to put an upper limit on the relative
mass difference of the two charge states of the
tau~\cite{ref:BelleMass}. They construct the
pseudomass~\cite{ref:pseudomass}, a kinematic quantity which, in
the absence of initial and final state radiation, displays a sharp
edge at the tau mass. By fitting the edge (see
Fig.~\ref{fig:MassDiff}), they obtain a mass $M_\tau =
1776.71\pm0.13\pm0.32$~MeV/$c^2$ and a relative mass difference
$|M_{\tau^+} - M_{\tau^-}|/M_\tau < 2.8 \times 10^{-4}$ at the
90\% confidence level (CL). The measurement of the relative mass
difference represents a significant improvement over the previous
limit on this quantity.

BaBar has made a preliminary measurement of the tau lifetime and
the relative lifetime difference of the two charge states, using
an 80~fb$^{-1}$ data sample. They measure a lifetime of $\tau_\tau
= 289.40 \pm 0.91 \pm 0.90$~fs and a relative difference,
$(\tau_{\tau^-} -\tau_{\tau^+})/(\tau_{\tau^-} + \tau_{\tau^+})$,
of $0.12 \pm 0.32$\% (the systematic error has still to be
evaluated). All of the measurements to date are consistent with
CPT invariance and with lepton universality.

\subsection{Searches for lepton flavor violation in tau decays}

Belle has recently published
results~\cite{ref:BelleLFV1}\cdash\cite{ref:BelleLFV2} on searches
for the LFV decays of the tau to a lepton plus two pseudoscalar
mesons ($\pi^\pm$ or K$^\pm$), including the cases where the
mesons are decay products of a K$^0_{\rm s}$, $\rho$, K$^*$ or
$\phi$. The most promising channel among these may be $\tau^- \to
\mu^-\rho^0$, where the minimal supersymmetric standard model
(MSSM) with small $\tan \beta$ could give a branching fraction as
large as $10^{-8}$. These decay modes may also be sensitive to new
physics models with heavy Dirac neutrinos and to models with
dimension-six effective fermionic operators that induce $\tau-\mu$
mixing. With a data sample of 158~fb$^{-1}$, Belle set 90\% CL
upper limits that vary from $1.6 \times 10^{-7}$ for $\tau^- \to
{\rm e}^-\pi^-{\rm K}^+$ to $8.0 \times 10^{-7}$ for $\tau^- \to
\mu^-{\rm K}^-{\rm K}^+$. Using 281~fb$^{-1}$ they obtain limits
of $5.6 \times 10^{-8}$ for e$^-$K$^0_{\rm s}$ and $4.9 \times
10^{-8}$ for $\mu^-$K$^0_{\rm s}$

Belle has also made a search~\cite{ref:BelleLFV3} for the decay
$\tau^- \to \mu^-\eta$ putting an upper limit at $3.4 \times
10^{-7}$ at 90\% CL. The Belle results improve on previous LEP and
CDF exclusion limits in the $\tan \beta$ versus $m_{\rm SUSY}$
plane.

BaBar has recently used a data sample of 233~fb$^{-1}$ to search
for the decay $\tau^- \to \mu^- \gamma$. In MSSM with the seesaw
mechanism, the branching fraction for this process is directly
related to the value of $\tan \beta$  and the mass of lightest
chargino. BaBar excludes this process~\cite{ref:BaBarmugamma} with
an upper limit on the branching fraction of $6.8 \times 10^{-8}$
at 90\% CL.

\section{Semihadronic Tau Decays}

The non-strange hadronic tau decays are dominated by vector and
axial hadronic states, with the vector part of the so-called
spectral function being particularly important as input to the
calculation of the muon magnetic moment anomaly, $a_\mu = (g_\mu
-2)/2$. The largest uncertainty in this calculation comes from the
lowest order hadronic correction to the photon propagator, whose
value is inferred from measurements of the cross section for the
process e$^+$e$^- \to {\rm {hadrons}}$ at low energies. Almost
three-quarters of this correction is due to the $\pi^+\pi^-$ final
state, dominated by the $\rho(770)^0$. The e$^+$e$^-$ annihilation
data can be related, by isospin invariance (conserved vector
current), to the decay $\tau^- \to \pi^-\pi^0\nu_\tau$ for which
much higher statistics measurements are available. In principle,
therefore, the tau decay data could provide more precise input for
$a_\mu$; however there remains a discrepancy between the
e$^+$e$^-$ and $\tau$ data which currently is not understood. A
recent review can be found in Ref.~\refcite{ref:murho}.

Studies of the strange spectral functions in hadronic tau decays
are the route to the world-best measurements of the CKM matrix
element $V_{\rm us}$ and the mass of the strange quark. All
measurements to date of the strange tau decays are limited by
statistics, and so the B-factory data have a very important role
to play. Already, the BaBar and Belle preliminary measurements of
the $\tau \to {\rm K}\pi\nu_\tau$ modes are more precise than the
previous world averages.

\subsection{$\tau^- \to \pi^-\pi^0\nu_\tau$ and muon $g-2$}

\begin{figure}[pb]
\centerline{\psfig{file=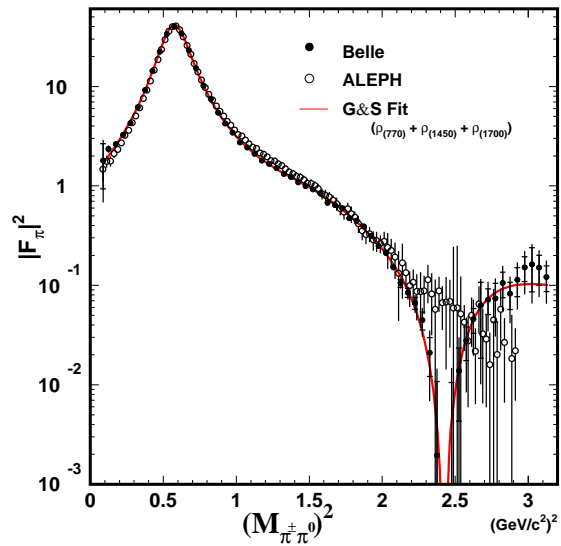,width=0.6\textwidth}}
\vspace*{8pt} \caption{Belle measurement of the pion form factor
(solid circles), fitted using a Gounaris-Sakurai model with all
parameters allowed to float. The open points show the ALEPH
results from Ref. \protect \refcite{ref:ALEPH2pi}.
\label{fig:Belle2pi}}
\end{figure}

Belle has used a sample of 72~fb$^{-1}$ to make a first study of
their $\pi^-\pi^0$ mass spectrum from $\tau$
decay~\cite{ref:Belle2pi}. They fit to the phenomenological model
of Gounaris and Sakurai~\cite{ref:GSmodel} which includes a set of
three interfering $\rho$ Breit-Wigner amplitudes, and they find
the need for a $\rho(1700)$ state in addition to the two
well-known $\rho(770)$ and $\rho(1450)$ resonances (see
Fig.~\ref{fig:Belle2pi}). Belle reports a branching fraction
BF$(\tau^- \to \pi^-\pi^0\nu_\tau) = (25.15\pm0.04\pm0.31)\%$. The
result for $a_\mu^{\pi\pi}$ is in agreement with previous tau
measurements from ALEPH~\cite{ref:ALEPH2pi} and
CLEO~\cite{ref:CLEO2pi} and in disagreement with the numbers
deduced from the e$^+$e$^-$ data~\cite{ref:murho}.

\subsection{The decay $\tau^- \to 3h^-2h^+\nu_\tau$}

\begin{figure}[htbp]
    \centering
    \subfigure[]{\label{fig:5pifigure1}
    \includegraphics[width=0.4\textwidth]{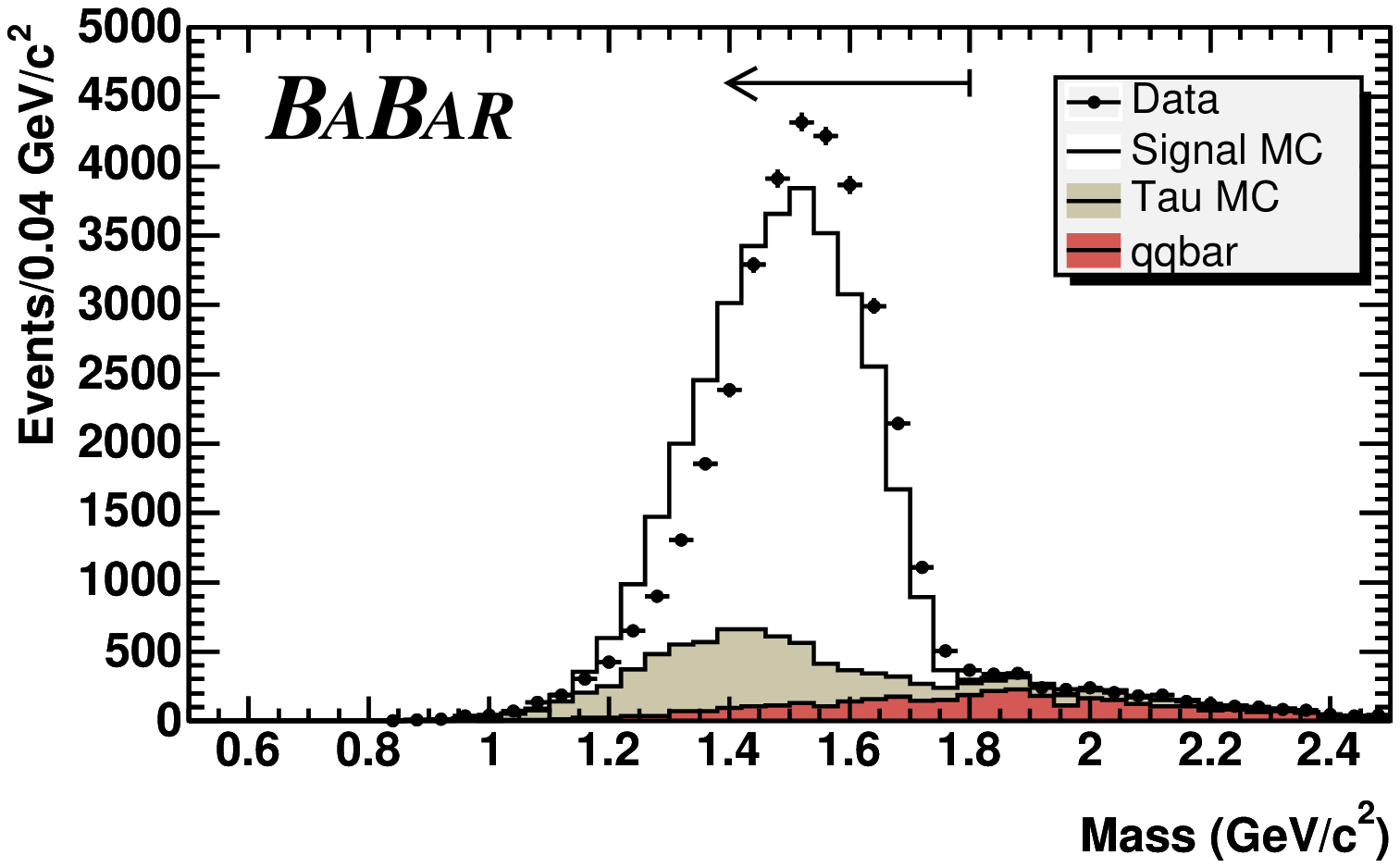}}
    \hspace{.2in}
    \subfigure[]{\label{fig:5pifigure2}
    \includegraphics[width=0.4\textwidth]{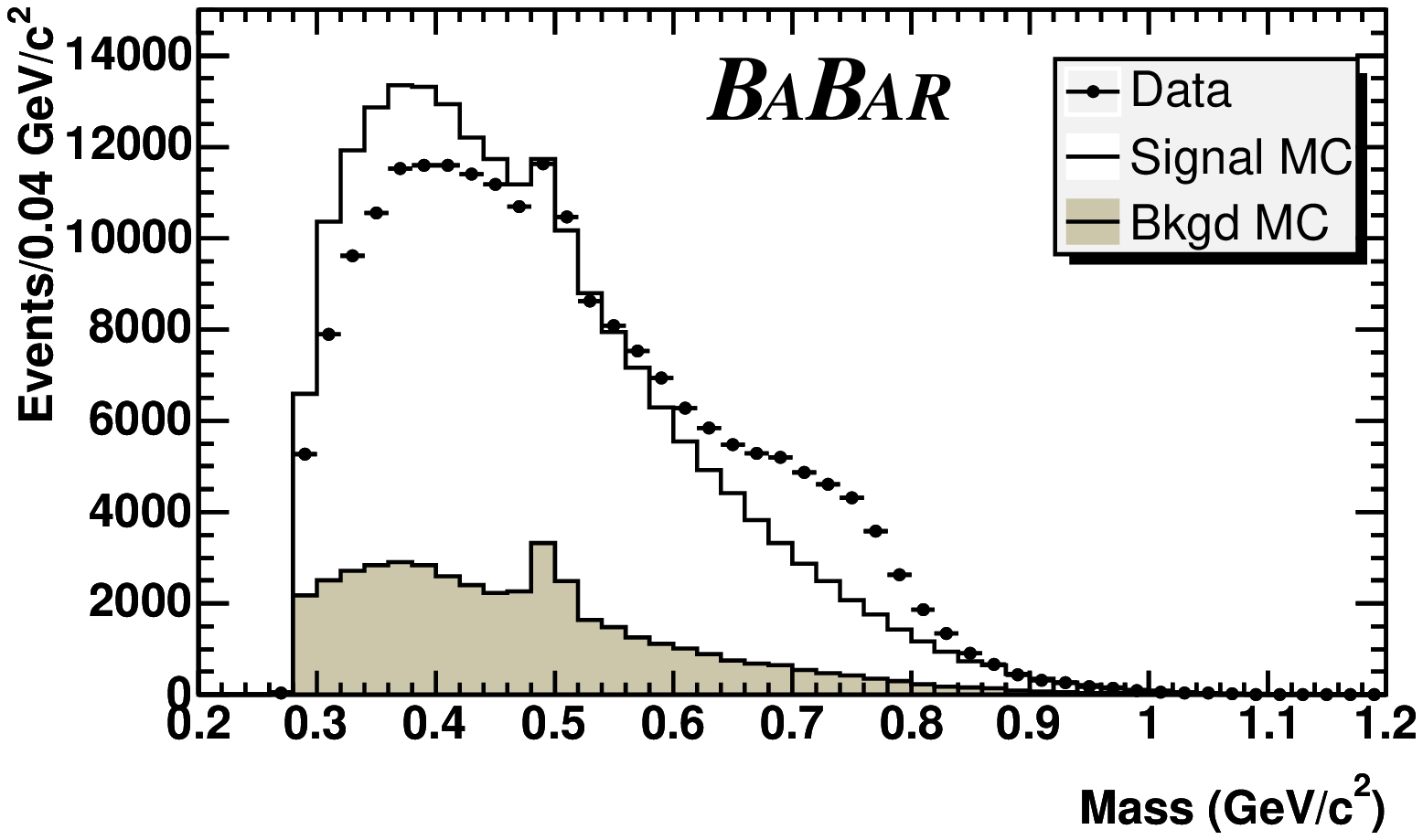}}
    \hspace{.2in}
    \subfigure[]{\label{fig:5pifigure3}
    \includegraphics[width=0.4\textwidth]{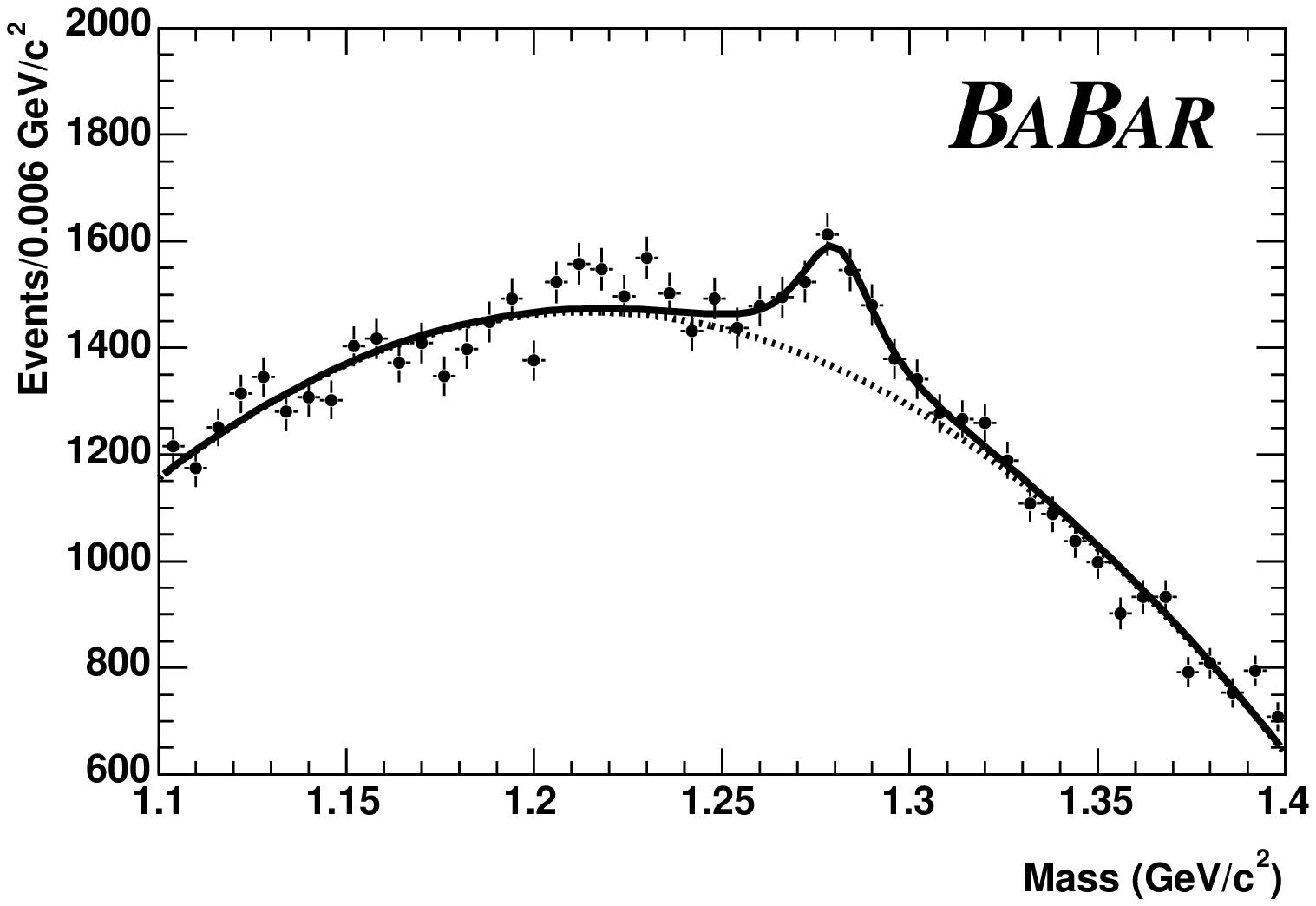}}
    \vspace*{8pt} \caption{Mass spectra from BaBar for the decay
    $\tau^- \to 3\pi^-2\pi^+\nu_\tau$, for (a) the $5\pi$ system;
    (b) inclusive $\pi^+\pi^-$ systems, and (c) $2\pi^+2\pi^-$ (with a fit
    including the f$_1(1270)$ meson. }
\end{figure}

In a recent BaBar analysis~\cite{ref:BaBar5pi}, a data sample of
232~fb$^{-1}$ has been used to study the process $\tau^- \to
3h^-2h^+\nu_\tau$. The branching fraction is measured as $(8.56
\pm 0.05 \pm 0.42) \times 10^{-4}$, representing a significant
improvement on the previous world average. The mass spectrum of
the hadronic system, with all tracks taken as pions, is shown in
Fig.~\ref{fig:5pifigure1}, where it is compared to the output of
the Tauola Monte Carlo model~\cite{ref:Tauola}.
Fig.~\ref{fig:5pifigure2} shows the inclusive $\pi^+\pi^-$ mass
spectrum, indicating a significant contribution from
$\rho(770)^0$. The $2\pi^+2\pi^-$ spectrum
(Fig.~\ref{fig:5pifigure3}) shows a signal from the f$_1(1285)$
meson, from which the branching fraction is measured to be
BF$(\tau^- \to {\rm f}_1{\rm h}^-\nu_\tau) = (3.9 \pm 0.7 \pm 0.5)
\times 10^{-4}$. These data will be used to improve the
implementation of these channels in the Tauola Monte Carlo.

\subsection{Searches for $\tau^- \to 3\pi^-2\pi^+2\pi^0\nu_\tau$ and
$\tau^- \to 4\pi^-3\pi^+(\pi^0)\nu_\tau$}

BaBar has looked for evidence for decays of the $\tau$ to seven or
more pions~\cite{ref:BaBar7pi}, which would be highly suppressed
by phase space. In principle, the rate of such decays would be
sensitive to the mass of the $\tau$ neutrino, but recent results
on the neutrino mass differences has somewhat reduced the interest
in this, since indications are that the neutrino masses are too
small to be probed by high-multiplicity $\tau$ decays.

Using 232~fb$^{-1}$ of data, BaBar has placed the following upper
limits on branching fractions at 90\% CL: $3.4\times 10^{-6}$ for
$\tau^- \to 3\pi^- 2\pi^+ 2\pi^0\nu_\tau$; $5.4\times 10^{-7}$ for
$\tau^- \to 2\omega(782) \pi^- \nu_\tau$; $4.3\times 10^{-7}$ for
$\tau^- \to 4\pi^- 3\pi^+ \nu_\tau$; and $2.5\times 10^{-7}$ for
$\tau^- \to 4\pi^- 3\pi^+ \pi^0\nu_\tau$.

\subsection{Strange hadronic tau decays}

BaBar has made a preliminary measurement of $\tau^- \to {\rm
K}^-\pi^0\nu_\tau$ with 124~fb$^{-1}$ of data, obtaining a
branching fraction of $(0.438 \pm 0.004 \pm 0.022)\%$. For the
complementary mode, $\tau^- \to {\rm K}^0_{\rm S}\pi^-\nu_\tau$,
Belle uses 351~fb$^{-1}$ to measure a branching fraction of
$(0.391 \pm 0.004 \pm 0.014)\%$. The two measurements are
consistent. In preliminary fits to their K$\pi$ mass spectrum,
Belle finds some evidence for the presence of the K$^*(1410)$
together with a scalar state (the so-called $\kappa$).

\section{Outlook}

Each of BaBar and Belle expects to accumulate samples of close to,
or greater than, 1000~fb$^{-1}$ by the end of 2008. Thus there is
the potential for much more $\tau$ physics in the coming years.
Limits on LFV could be reduced down to $\sim 10^{-8}$ for some
channels, with even the possibility of the emergence of evidence
for new physics. In semihadronic decays, measurements of the
strange spectral functions should provide an improved measurement
of the CKM matrix element $V_{\rm us}$ and perhaps also of the
mass of the strange quark. Measurements of the non-strange
channels will include the important vector spectral functions,
which should continue to aid understanding of the
non-perturbative, hadronic contribution to the anomalous magnetic
moment of the muon.

\section*{Acknowledgments}

Many thanks go to the BaBar and Belle Collaborations, whose work I
have reported. I would also thank the organizers of Charm 2006 for
what was a most enjoyable and productive workshop.

%
%


\end{document}